# From Divergence to Consensus: Evaluating the Role of Large Language Models in Facilitating Agreement through Adaptive Strategies

Loukas Triantafyllopoulos [a] [*] and Dimitris Kalles [a]

[a] School of Science and Technology, Hellenic Open University, Patras, Greece.

[*] Corresponding Author:

Loukas Triantafyllopoulos, PhD Candidate, School of Science and Technology, Hellenic Open University, 18 Aristotelous Stz, 263 35 Patras, Greece.

(e) triantafillopoulos.loukas@ac.eap.gr (t) 0030 6934222370

**Abstract**

Achieving consensus in group decision-making often involves overcoming significant challenges, particularly in reconciling diverse perspectives and mitigating biases that hinder agreement. Traditional methods relying on human facilitators are often constrained by scalability and efficiency, especially in large-scale, fast-paced discussions. To address these challenges, this study proposes a novel framework employing large language models (LLMs) as automated facilitators within a custom-built multi-user chat system. Leveraging cosine similarity as a core metric, this approach evaluates the ability of three state-of-the-art LLMs—ChatGPT 4.0, Mistral Large 2, and AI21 Jamba-Instruct—to synthesize consensus proposals that align with participants' viewpoints. Unlike conventional techniques, the system integrates adaptive facilitation strategies, including clarifying misunderstandings, summarizing discussions, and proposing compromises, enabling the LLMs to iteratively refine consensus proposals based on user feedback. Experimental results demonstrate the superiority of ChatGPT 4.0, which achieves higher alignment with participant opinions, requiring fewer iterations to reach consensus compared to its counterparts. Moreover, analysis reveals the nuanced performance of the models across various sustainability-focused discussion topics, such as climate action, quality education, good health and well-being, and access to clean water and sanitation. These findings highlight the transformative potential of LLM-driven facilitation for improving collective decision-making processes and underscore the importance of advancing evaluation metrics and cross-cultural adaptability in future research.

# Introduction

The concept of "achieving consensus" is often misinterpreted as complete unanimity among members of a group [1]. In fact, consensus does not require absolute agreement, but the acceptance of a proposal that is considered satisfactory by all interested parties, taking into account the constraints that exist [2]. Nevertheless, achieving this kind of agreement is a complex challenge, especially when participants have conflicting views [3, 4]. Differences in opinion often arise from personal beliefs and priorities [5], while biases like attachment to initial positions [6, 7] and confirmation bias [8] can hinder consensus by limiting openness to alternative perspectives.

Given the importance of this issue in collective decision-making, particularly on matters of significant concern to humanity, various methods have been employed over time to facilitate the process. One common approach is the inclusion of a facilitator, often an expert or knowledgeable individual on the topic under discussion [9, 10]. The facilitator's role is to guide and structure the dialogue, helping participants navigate differing viewpoints, identify common ground, encourage critical thinking, and propose mutually acceptable ideas to reach consensus [11]. While this practice has proven effective in many cases, it has shown limitations when addressing particularly complex or contentious issues, especially in the context of large-scale and fast-paced online discussions [12, 13].

On this issue, the research community has increasingly focused on automating the role of the facilitator [14], with the use of artificial intelligence (AI) playing a pivotal role in advancing this effort in recent years [12]. Notably, a growing body of research highlights the potential superiority of AI-driven facilitators over their human counterparts [13, 15]. This trend has further evolved with the adoption of large language models (LLMs), which are increasingly being recognized as effective tools for facilitating complex discussions and consensus-building processes [16].

LLMs possess a unique ability to process and synthesize vast amounts of information [17]. By analyzing diverse perspectives, identifying common themes, and proposing neutral, balanced statements, they can serve as powerful tools for fostering consensus [18]. Additionally, acting as unbiased facilitators, these models can synthesize varied viewpoints, ensuring all group members have an equitable voice in the conversation [19].

To evaluate LLMs' success in achieving consensus, measures like cosine similarity are particularly useful [3, 20, 21]. Cosine similarity is effective in ranking alternatives and assessing how closely participants' views align during the consensus process [22], often providing more accurate results than Euclidean distance in such scenarios [23]. Additionally, cosine similarity allows for the assessment of alignment in high-dimensional spaces [24], which is common in LLM-based tasks where the complexity of the data often extends beyond simple scalar comparison.

In this framework, the current study aims to evaluate the effectiveness of three popular LLMs in facilitating consensus among individuals on critical issues. To accomplish this, a chat-based environment was developed from scratch, enabling synchronous communication between participating users and following a predefined discussion scenario where the LLM served as the facilitator. Especially, its role was to synthesize consensus proposals based on user feedback, employing, in cases of disagreement, four well-known strategies from the literature for achieving consensus. After that, the study used cosine similarity to measure the alignment between users' initial positions and the final consensus, providing a precise metric to evaluate the effectiveness of these strategies and the overall ability of the LLMs to foster agreement.

The key contributions of this paper can be outlined as follows:

1. First, the creation of a custom platform for synchronous communication, where the LLM serves as a facilitator, providing a novel framework for studying consensus-building interactions.
2. Second, the introduction of cosine similarity as a precise metric to quantify alignment between users' initial positions and the final consensus, offering an objective way to assess the effectiveness of consensus-building strategies.
3. Finally, a systematic comparison of three popular LLMs, providing valuable insights into their strengths, limitations, and potential for facilitating collaborative decision-making.

The organization of this paper is structured as follows: Section 2 reviews relevant literature and outlines the research questions that guide this study. Section 3 details the methodology employed, including the consensus-building framework and tools used, and explains the selection criteria for the sample in the pilot test. Section 4 presents the empirical findings. Section 5 discusses these results, explores their implications, and addresses the limitations of the study. Finally, Section 6 offers concluding remarks and highlights the key insights derived from the research.

## 2. Background

Within the context of the relevant literature, we focus our attention on two critical issues. The first concerns the role of artificial intelligence (AI) in facilitating the process of consensus building; the second addresses the various approaches used to measure AI's effectiveness in this area. After reviewing the pertinent information, we present the approach of the current study as an advancement over existing literature, highlighting the gaps it aims to fill and its potential contributions to the scientific community.

### *2.1. The Role of Artificial Intelligence in Facilitating Consensus Building*

A review of the recent literature reveals numerous research efforts aimed at evaluating AI as a reliable tool for mediating and facilitating consensus building. For this purpose, scholars often compare the outcomes of two scenarios: one with AI assistance and one without it throughout the entire process.

One popular approach discussed in the literature involves regularly integrating conversational agents into discussion platforms such as D-agree [15, 25-28]. This online text-based platform allows participants to exchange messages with each other and with conversational AI. The AI is based on the Issue-Based Information System (IBIS) framework [29], which helps guide participants toward consensus by deconstructing dialogues into issues, ideas, pros, and cons. Additionally, these agents can summarize data, visualize discussion structures, and enhance participation with adaptive messages. Utilizing this framework, Haqbeen et al. conducted studies on achieving consensus regarding sustainability issues in Kabul, Afghanistan, and addressing solidarity in the Ukraine conflict [25, 26]. Similarly, Sahab et al. [28] noted the effectiveness of AI agents in facilitating consensus in diverse environments across five Afghan cities, while Ito et al. [15] assessed the social influence of this platform in Japan. Indeed, Sahab et al. [30] later compared the results between these two countries to evaluate the effectiveness of AI agents across different cultural settings. Finally, Hadifi et al. [27] explored how conversational agents could act as facilitators to bolster women's participation in online debates in culturally conservative settings.

Despite the popularity of the D-agree platform and the use of IBIS for guiding the behavior of AI agents, some researchers consider the use of IBIS problematic in facilitating consensus achievement. Dong [31] cites two critical issues. Regarding the discussion environment, he argues that it is not inclusive, as various characteristics of participants are ignored, leading to insufficient discussions. Additionally, he contends that the facilitation messages generated by IBIS-based AI agents are not sufficiently natural, resulting in reduced participant engagement and a lack of flow in topic discussions. As an ideal alternative, Dong et al. [13] propose the use of LLMs for guiding AI agents, presenting a series of additional functionalities that these models offer. They strengthen their position by comparing the results of the two approaches in their publication, distinguishing LLMs (tested using GPT-3.5) as more effective for assisting the consensus process. In a related development, Nomura et al. [32] have historically integrated LLMs with the IBIS framework to enhance brainstorming interactions. By employing GPT-3.5-turbo, they assigned AI agents roles that facilitated more dynamic and inclusive discussions, addressing potential concerns about engagement and the natural flow of dialogue. This approach leveraged the advanced capabilities of LLMs to improve the effectiveness of consensus-building processes, demonstrating the benefits of merging structured facilitation methods with the adaptability of modern AI technologies.

The superiority of LLMs as tools for facilitating consensus is recognized by other researchers in various ways. For example, Ding and Ito [3] developed the Self-Agreement framework, utilizing GPT-3 to autonomously generate and evaluate diverse opinions, thereby facilitating consensus without human input. Taking a different approach, Song et al. [33] demonstrated how multi-agent AI systems, powered by LLMs, can significantly influence social dynamics and steer opinions towards consensus by increasing social pressure among participants. Additionally, Tessler et al. [34] introduced the 'Habermas Machine,' an AI designed to act as a mediator in discussions on divisive political topics. Their AI was trained to produce group statements that not only garnered wide agreement among UK groups discussing issues like Brexit and immigration but also did so more effectively than human mediators. The 'Habermas Machine' helped in creating statements that were clearer, more logical, and informative, ensuring minority perspectives were not alienated, thus demonstrating another powerful application of LLMs in consensus-building within diverse groups.

Building on the capabilities of LLMs to facilitate consensus building, Govers et al. [12] focused on the specific strategies that AI agents must employ to effectively mediate online debates. Their research utilized the Thomas-Kilmann Conflict Mode Instrument (TKI) to implement various conflict resolution strategies through LLMs. The study particularly emphasized high-cooperativeness strategies, which were found most effective at depolarizing discussions and enhancing the perception of achievable consensus among participants. This approach demonstrated how tailored AI-driven strategies could significantly influence online interactions, guiding diverse groups toward agreement and more constructive dialogue.

*2.2. Measuring the effectiveness of AI in Consensus-Building*

As noted in the previous section, various approaches illustrate how AI can be a useful and necessary tool for facilitating the consensus-building process. However, a subsequent step involves assessing the extent to which AI achieves this goal. In the literature, numerous metrics and methodologies have been proposed to measure AI's impact on consensus formation, ranging from semantic similarity scores to perceptual and qualitative assessments. This diversity reflects the complexity of consensus-building, which is shaped by both objective factors, such as agreement metrics, and subjective elements, such as participant satisfaction and perceived fairness.

One prominent line of research focuses on using semantic models to compare participant views and gauge overall agreement. Specifically, Ding and Ito [3] introduce a scoring function that evaluates candidate agreements generated by improved language models, relying on semantic similarity scores—such as cosine similarity computed with BERT embeddings—to determine how well different viewpoints are unified. This logic is also applied by Pérez et al. [35], who draw on soft consensus measures and fuzzy logic to quantify partial agreements, using indices like Euclidean, Manhattan, and cosine distances, along with adaptive thresholds.

Another approach examines the broader social context influenced by AI. Song et al. [33] use a 6-point Likert scale to measure opinion shifts and polarization before and after AI-mediated discussions, shedding light on the social forces that either foster or hinder consensus. Likewise, Hadfi and Ito [36] adopt "Augmented Democratic Deliberation," using feedback loops in knowledge graphs to assess the depth and quality of group deliberations.

Additionally, exploring the subjective experiences of participants sheds further light on how AI can affect consensus outcomes. Govers et al. [12] introduce a Perceived Consensus Score that allows participants to indicate how realistic they believe potential agreements are, potentially revealing unspoken reservations or hidden divergences. This metric is complemented by qualitative insights from Pham et al. [5] and de Rooij et al. [37], who conduct interviews to capture richer details about participants' perceptions of AI's role in shaping group dynamics and decision-making.

Finally, integrating both quantitative and qualitative data provides a comprehensive view of how AI can facilitate consensus. Kim et al. [38] and Sahab et al. [39] show that AI-facilitated platforms not only lead to measurable shifts in opinions but also improve perceptions of fairness and inclusivity. This dual impact underscores AI's potential to enhance the structural efficiency of consensus-building—via semantic modeling and feedback loops—as well as the human experience of collaboration. By fostering more productive, transparent, and balanced discussions, AI interventions increase the likelihood of durable and broadly accepted agreements.

## 3. A layered path to opinion convergence

Building on the relevant literature on using AI to facilitate consensus-building, and particularly the potential of LLMs for this purpose, we developed a systematic approach to both the process and the measurement of its effectiveness. Specifically, we designed a web application from the ground up to enable structured communication and message exchange among participants, as well as interactions with the LLM models under examination: ChatGPT 4.0, Mistral Large 2, and AI21 Jamba. These models were integrated into an organized discussion framework that incorporated specific consensus-facilitating strategies, allowing each model to select appropriate actions based on conversation history and user feedback. The discussions were analyzed to evaluate the extent to which participants reached a consensus proposal aligned with their initial positions. The following section outlines the employed approach and its key components.

*3.1 System Architecture and Workflow*

The system architecture of the web application was designed to facilitate efficient interaction between participants and the LLMs, supporting real-time consensus-building processes. Figure 1 illustrates the consensus formation workflow, which begins with participants responding to a question posed during the discussion and expressing their opinions on the topic. The LLM synthesizes these opinions to formulate a proposal aimed at achieving consensus among the participants.

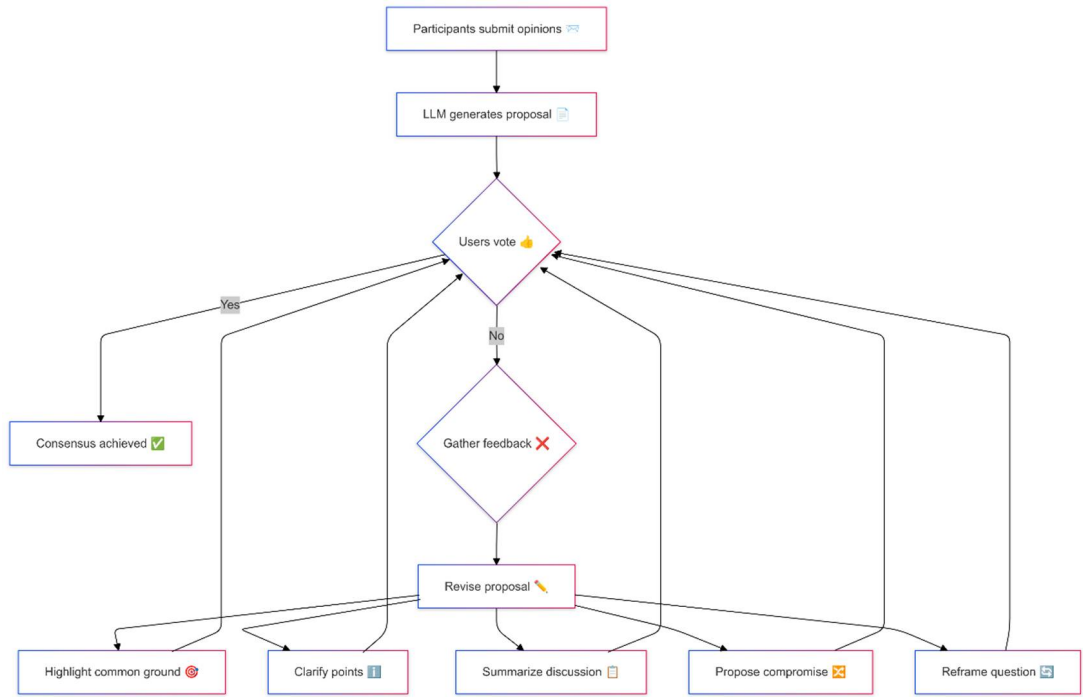

**Fig. 1.** Consensus Formation Workflow

Participants then indicate their acceptance or rejection of the consensus proposal. In cases of rejection, the LLM collects feedback to understand the reasons, while participants who agree await the feedback process. The LLM considers the conversation history and user feedback to select one of five predefined strategies. Each strategy guides the LLM in creating a revised proposal using tailored prompts. This iterative process continues until a consensus is reached or the discussion ends, with the LLM adapting its strategy based on the conversation dynamics.

The five strategies integrated into the workflow are summarized in Table 1. These strategies, derived from a literature review, aim to promote understanding, summarize key points, highlight agreements, propose compromises, and reframe questions when necessary. For instance, Bosch et al. [40] and Dworetzky et al. [41] emphasize the importance of the 'ClarifyUnderstanding' strategy for eliminating ambiguities, while Chosokabe et al. [42] highlight the 'SummarizeDiscussion' strategy for condensing complex conversations into their essential elements. Similarly, Nieto-Romero et al. [43]

stress the value of the 'HighlightCommonGround' approach, which effectively identifies shared agreements among diverse perspectives. Complementing these strategies is the 'ProposeCompromise' method, emphasized by Opricovic [44] and Basuki [45], which seeks to offer balanced solutions that accommodate differing viewpoints. Additionally, Kettenburg et al. [46] advocate for the 'ReframeQuestion' strategy, which involves skillfully rephrasing or redirecting research questions to foster new insights and facilitate consensus.

**Table 1**

Available Strategies for Selection by the LLM to Achieve Overall Consensus

| Strategy | Prompt |
| --- | --- |
| ClarifyUnderstanding | Provide additional explanations, definitions, or examples to eliminate misunderstandings or ambiguities related to the research question or discussion points. |
| SummarizeDiscussion | Provide a concise summary of the discussion, highlighting key points of agreement and disagreement. |
| HighlightCommonGround | Identify and emphasize points of agreement among participants. |
| ProposeCompromise | Suggest potential compromises or middle-ground solutions. |
| ReframeQuestion | Rephrase or adjust the focus of the research question to make it more agreeable or clearer. |

To support this architecture, we utilized several technologies, including HTML, JavaScript, PHP, CSS, and AJAX, to ensure the application's user-friendliness and compatibility across devices. Real-time interaction with the LLMs was facilitated through an open-source server environment built using Node.js. Conversation history, critical for sustained discussions, was stored in a specially configured MySQL database. Finally, API technology was employed to integrate the LLMs into the platform, with necessary code modifications made to align with the previous described workflow in Figure 1.

## 3.2. The characteristics of the evaluated LLMs

The three LLMs selected in this research are distinguished by their popularity and promising results in addressing complex tasks. Developed by three different companies, each model possesses unique features while sharing common attributes.

### 3.2.1. The ChatGPT Model

ChatGPT, developed by OpenAI, has rapidly gained popularity due to its powerful capabilities [47]. Version 4.0, released in March 2023, has been extensively tested across various tasks by the research community, including its effectiveness in facilitating consensus among participants [33].

This success is partly due to its architecture, which follows the Transformer model [48; see Figure 1] — an approach that excels at processing input sequences and generating outputs by predicting the next word in a sentence based on context [49]. Unlike previous recurrent neural network models, which process text sequentially from left to right, the Transformer can simultaneously consider all words in a sentence, even those at the ends of the input. It achieves this by employing multiple levels of self-attention mechanisms, which calculate a weighted sum of the input sequence. These weights are determined based on the similarity between each input token and the rest of the sequence [50].

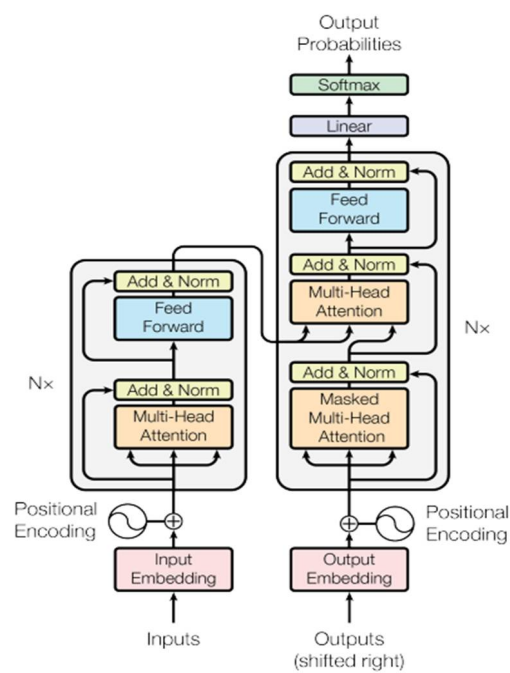

**Fig. 2**. The Transformer - model architecture (Vaswani, 2017)

In addition, ChatGPT introduced some unprecedented advancements. Specifically, a unique training approach, based on unsupervised learning, was employed to autonomously process vast amounts of internet data. This approach involved incorporating approximately 1.76 trillion parameters [51] to capture intricate patterns and connections in textual information. Furthermore, reinforcement learning with human feedback was applied in a subsequent step to enhance the model's responses, refining its ability to generate contextually appropriate and accurate outputs [52].

### 3.2.2. *The Mistral Large 2 model*

Mistral Large 2 is an LLM developed by Mistral AI and released in July 2024, significantly later than ChatGPT [53]. Unlike the previous mentioned LLM, this model was trained predominantly on French texts rather than English and utilizes a smaller number of parameters, approximately 123 billion [54]. Despite this difference, performance measurements indicate that it is an equally reliable choice, offering competitive results in general knowledge and reasoning tasks [55]. Furthermore, Mistral Large 2 supports dozens of foreign languages and more than 80 programming languages (e.g., Python, Java, C, C++, JavaScript, Bash), making it highly capable in multilingual text processing and code generation [56, 57].

Mistral Large 2 also builds upon the traditional Transformer architecture [55], introducing several improvements aimed at reducing computational complexity without significantly compromising the quality of its outputs. One notable enhancement is the Sliding Window Attention mechanism, which limits attention to a specific window of tokens. Additionally, the model incorporates mechanisms such as the Key-Value Cache with Rolling Buffer, enabling the storage and reuse of precomputed values during text generation. This reduces redundant calculations and improves the model's speed and efficiency [58]. Furthermore, as an open-source LLM, Mistral Large 2 allows the research community to make modifications, fostering further advancements and optimizations [53].

### 3.2.3. *The AI21 Jamba-Instruct Model*

The third and final model evaluated in this study is AI21's Jamba-Instruct, which was released in early May 2024 [59]. This particular LLM has been trained on a relatively smaller number of parameters, approximately 52 billion [60]. Despite this, it achieves remarkable results across a series of performance indicators, making it a standout choice among open-source LLMs [61].

The architecture of the Jamba model series by AI21 Labs is particularly noteworthy. It employs an innovative hybrid approach that extends beyond the traditional Transformer architecture by integrating Mamba layers. These layers enhance the processing of long-distance relationships in large texts while reducing memory requirements. Additionally, the inclusion of Mixture of Experts (MoE) enables the use of multiple experts per layer, significantly increasing the model's capacity without a proportional increase in computational cost. It's able to handle content lengths of up to 256,000 tokens, with a KV cache memory footprint of only 4GB—substantially lower than the 32GB required by comparable LLMs. This efficiency enables Jamba to achieve top-tier performance in benchmarks such as HellaSwag and MMLU, all while being up to three times more efficient than other models of similar size [60, 61].

### *3.3. Calculating text similarity as a surrogate for consensus*

A common and widely used method for comparing text similarity is the cosine similarity index [62]. This index is efficient in computation [63] and its results are easy to interpret [64]. It quantifies the cosine of the angle between two non-zero vectors in a multidimensional space [65]. Specifically, its calculation formula is defined by the following equation [66]:

$$\text{Cosine Similarity：} \frac{A*B}{\|A\|\,\|B\|} \quad (1)$$

*where*

- A and B A represent the user opinion and the consensus sentence provided by LLM, respectively, as the two vectors.
- A * B represents the dot product of the two vectors
- $\|A\|$ and $\|B\|$ represent the magnitudes (or norms) of the vectors A and B.

The resolution of this relationship involves four distinct steps. First, the texts under consideration are converted into vector representations, followed by the calculation of the dot product between them. Next, the norms of these vectors are computed. Finally, the cosine similarity is determined by dividing the dot product by the product of the magnitudes of the two vectors.

In this study, this process was applied to analyze the Cosine Similarity between the initial opinions of the participants and the final proposal that was ultimately accepted, in the effort to achieve consensus. Additionally, it was also used to assess the differences between intermediate states, namely, the opinions of participants regarding consensus proposals that were not accepted. The study required participants to provide feedback on the reasons for non-acceptance of these proposals. This analysis helped to determine the extent to which participants maintained their original positions throughout the discussion or modified their stances by accepting proposals that did not align with their initial positions.

In the first step of this methodology, the examined texts were converted into vector representations using the Universal Sentence Encoder (USE) available from TensorFlow Hub. This choice is advantageous compared to other methods such as TF-IDF (Term Frequency-Inverse Document Frequency) and Word Embeddings, as it is capable of capturing the semantic content of entire sentences or paragraphs [67]. The vectorization process was implemented using Python code on the Google Colab platform.

Next, we proceeded to calculate the Dot product (numerator of equation 1), which concerns the sum of the vector products between each user's opinions and the consensus proposals from the LLMs. More specifically, we have the following equation (2):

$$A * B = \sum_{i=0}^{n} A_i * B_i \quad (2)$$

*where*

- $A_i$ and $B_i$ are the elements of vectors A and B, respectively, and
- n is the dimension of the vectors

Then, we calculated the magnitudes (norms) of the vectors by taking the square root of the sum of the squares of their elements. These magnitudes represent the length of each vector in the multidimensional space. In this context, the following holds true:

$$\|A\| = \sqrt{\sum_{i=1}^{n} A_i^2} \quad \text{and} \quad \|B\| = \sqrt{\sum_{i=1}^{n} B_i^2} \quad (3)$$

Finally, we calculated the cosine similarity by dividing the dot product, as detailed in relation (2), by the product of the magnitudes of the two vectors from relation (3). This index ranges between -1 and 1, where:

- A cosine similarity of 1 indicates that the vectors are identical, meaning there is complete agreement between the user's proposal and the consensus proposal.
- A cosine similarity of 0 indicates that the vectors are orthogonal, suggesting no alignment between the user's proposal and the consensus proposal generated by the LLM.
- A cosine similarity of -1 indicates that the vectors are diametrically opposed, meaning the user's proposal and the consensus proposal from the LLM are completely opposite.

## 4. Experimental Setup

The study aimed to evaluate the AI-assisted application by focusing on a group with diverse views on sustainability, making students the ideal choice for participation [68, 69]. In August 2024, the Hellenic Open University launched a call for volunteers, specifically targeting Informatics students from all years through its official website. The announcement provided a straightforward application process with clear instructions, a consent form, and a video guide to ensure participants could easily navigate the application. To uphold ethical standards and protect privacy, the University's Ethics and Research Committee approved the study, requiring participants only to create a personal username and password, with anonymity fully safeguarded throughout.

Additionally, the implementation of the sessions between participants and the use of the three LLMs via API required resources, which were provided by both OpenAI and Amazon. Specifically, OpenAI offered free API access through the "OpenAI Researcher Access Program," while the other two models were made available through Amazon Bedrock at no cost as part of the Open Cloud for Research Environments (OCRE) initiative.

In total, thirty students expressed interest in participating in the research following the University's invitation. These students, using the specially designed management platform, organized sessions among themselves, sending invitations, selecting their questions, and scheduling the day and time for each session. As a result, between August 24 and September 30, 2024, a total of 75 sessions were conducted, with each student participating in at least one session. The sessions were evenly distributed across the three LLM models under study, with the platform occasionally adjusting the question availability to ensure a balanced evaluation of each model.

A total of six questions were selected for discussion, covering several sustainable development goals, such as good health and well-being (Goal 3) and quality education (Goal 4). The set of questions used is listed in Table 2 below.

**Table 2**
Questions for Discussion and Consensus Building

| # | Question |
|---|---|
| 1 | Should patients with a healthy lifestyle be prioritized in healthcare provision compared to those who choose a lifestyle that increases the risk of serious conditions? |
| 2 | Should international collaborations be strengthened to address pandemics like COVID-19, or would it be better to focus on national strategies to protect public health? |
| 3 | Should priority be given to the integration of digital technologies in education to better prepare students for the modern age, or should we focus more on strengthening basic skills in literacy and numeracy? |
| 4 | Should educational policy place greater emphasis on personalized learning and support for students with different learning needs, or is it more important to maintain uniform standards and approaches for all students? |
| 5 | Should global cooperation be strengthened to improve access to clean drinking water and sanitation services, or is it better to tailor policies to the specific needs of each region? |
| 6 | Should economically disadvantaged countries adopt stricter policies to reduce carbon dioxide emissions, or should greater emphasis be placed on adaptation and the resilience of societies to climate change? |

## 5. Results and Analysis

This section presents various analyses using cosine similarity calculations to compare the three LLMs in terms of their ability to facilitate the process of users reaching consensus during a controversial discussion. The analysis was performed at two levels: first, by examining the distance between the participants' initial and final positions, and second, through a more detailed graphical evaluation of the intermediate stages that evolved, ultimately leading to an acceptable consensus proposal.

*5.1. Evaluation of the distance between users' initial and final positions*

Regarding the first part of the results, Tables 3 and 4 are presented. Table 3 initially analyzes the degree of similarity between the users' initial and final positions at the model level, using the average cosine similarity as a metric. As shown, ChatGPT 4.0 achieves the highest value of 0.701, compared to 0.6126 for A21 Jamba and a relatively lower value of 0.5807 for Mistral Large 2. For the total of 124 calculated values, the overall average cosine similarity index was 0.6381.

**Table 3**
Average Cosine Similarity per examined LLM

| LLM | No of Occasions | Average Cosine Similarity |
| --- | --- | --- |
| ChatGPT 4.0 | 50 | 0,701 |
| Mistral Large 2 | 40 | 0,5807 |
| AI21 Jamba | 34 | 0,6126 |
| **Total** | **124** | **0,6381** |

Next, Table 4 examines the extent to which the LLMs facilitated consensus, this time considering the discussion topics, which were related to four of the 17 Sustainable Development Goals (SDGs). As shown, higher average cosine similarity indices were observed in discussions on good health and well-being, as well as on quality education—topics for which our study collected a larger sample. Specifically, for the two questions related to SDG 3 (Good Health and Well-Being)—'Should patients with a healthy lifestyle be prioritized in healthcare provision compared to those who choose a lifestyle that increases the risk of serious conditions?' and 'Should international collaborations be strengthened to address pandemics like COVID-19, or would it be better to focus on national strategies to protect public health?'—the average cosine similarity index was 0.658. Similarly, for the topics associated with SDG 4 and especially Quality Education — 'Should priority be given to the integration of digital technologies in education to better prepare students for the modern age, or should we focus more on strengthening basic skills in literacy and numeracy?' and 'Should educational policy place greater emphasis on personalized learning and support for students with different learning needs, or is it more important to maintain uniform standards and approaches for all students?' — the resulting average cosine similarity index reached 0.650. In contrast, a lower average cosine similarity index was observed for discussions on climate change (0.599), while the lowest index, 0.552, was recorded for the question addressing the global issue of clean water and sanitation.

**Table 4**

Average Cosine Similarity per Sustainable Task and LLM

| Sustainable Task | LLM | No of Occasions | Average Cosine Similarity |
|---|---|---|---|
| Good Health and Well-Being | ChatGPT 4.0 | 22 | 0,70499 |
|  | Mistral Large 2 | 16 | 0,5962 |
|  | AI21 Jamba | 14 | 0,657 |
|  | **Total** | **52** | **0.658** |
| Climate action | ChatGPT 4.0 | 2 | 0,8487 |
|  | Mistral Large 2 | 2 | 0,557 |
|  | AI21 Jamba | 4 | 0,4966 |
|  | **Total** | **8** | **0,599** |
| Quality Education | ChatGPT 4.0 | 20 | 0,7069 |
|  | Mistral Large 2 | 16 | 0,5921 |
|  | AI21 Jamba | 12 | 0,6355 |
|  | **Total** | **48** | **0.650** |
| Clean water and sanitation | ChatGPT 4.0 | 6 | 0,6211 |
|  | Mistral Large 2 | 6 | 0,5166 |
|  | AI21 Jamba | 4 | 0,5033 |
|  | **Total** | **16** | **0.552** |

At the LLM level, ChatGPT 4.0 demonstrates clear superiority in handling different discussion topics, consistently leading participants to consensus proposals that align with their initial views (see Table 4). For example, when addressing climate change, ChatGPT 4.0 achieves a significantly high similarity score of 0.8487, far surpassing the performance of Mistral Large 2, which scores 0.557, and AI21 Jamba, which scores 0.4966. Similarly, in the area of good health and well-being, ChatGPT 4.0 maintains its lead with a strong similarity index of 0.70499, outperforming AI21 Jamba (0.657) and Mistral Large 2 (0.5962). This trend continues with the goal of quality education, where ChatGPT 4.0 achieves an index of 0.7069, once again surpassing AI21 Jamba and Mistral Large 2. Even in the domain of clean water and sanitation, where the differences are narrower, ChatGPT 4.0 maintains its dominance with a similarity score of 0.6211, outperforming both AI21 Jamba and Mistral Large 2, which have scores of 0.5033 and 0.5166, respectively.

*5.2. Assessment of the intermediate state until users reach consensus*

To gain a better understanding of the consistency of the examined LLMs in generating proposals that align with the participants' feedback and the conversation history throughout the interaction until consensus is reached, graphs 3 to 5 are presented below.

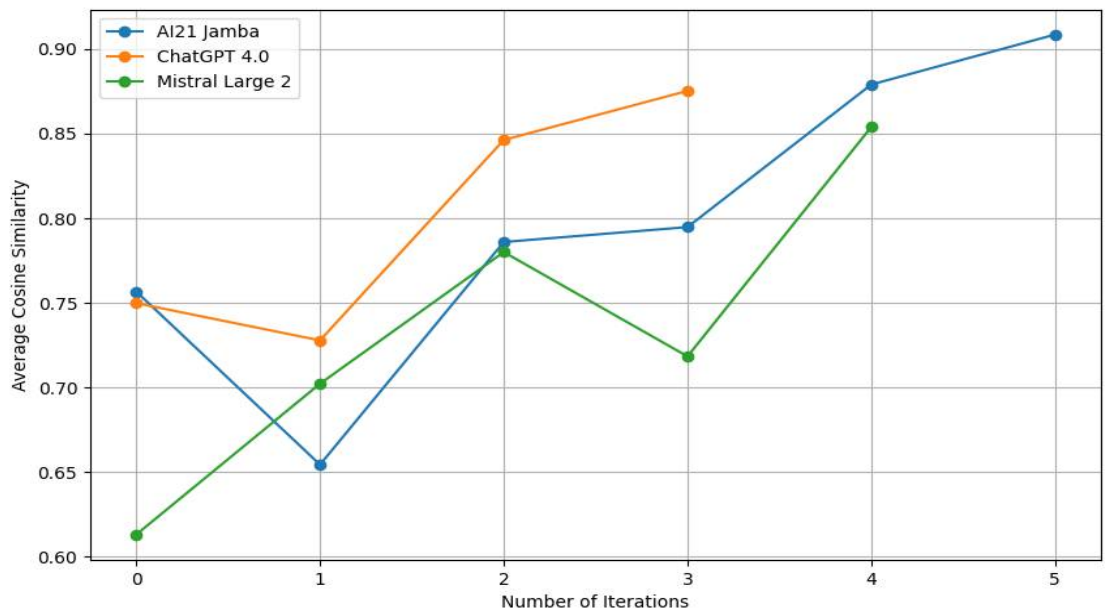

**Fig. 3**. Average Cosine Similarity across Iterations

In Graph 3, the variation of the average cosine similarity per LLM is observed as the discussion progresses, particularly in cases where no acceptable consensus proposal exists among the participants. As shown in the graph, even in the initial phase of interaction between the AI and users, ChatGPT 4.0 demonstrates a cosine similarity index close to 0.75. This value represents the similarity between the consensus proposal it generates and the initial opinions of the participants. This index increases significantly, particularly after the second and third iterations, as the model applies effective strategies to assist the discussion. Notably, in all discussions involving this LLM, no more than three iterations were required for the sample to reach a consensus. Similarly, AI21 Jamba achieves a high initial cosine similarity of 0.75. However, its behavior diverges as the process continues. According to Graph 3, its index drops significantly to 0.65 after the first round of user feedback but subsequently aligns with a more coherent integration of user opinions and feedback in later iterations. Finally, Mistral Large 2 starts with a notably lower initial cosine similarity index, slightly above 0.60, during its initial processing of users' opinions. However, it improves with subsequent feedback, ultimately reaching its highest value of 0.85 after the fourth iteration, when the process of selecting strategies to facilitate unanimity is repeated.

Next, Graph 4 illustrates the distribution of cases across different phases of the consensus-building process. In this graph, ChatGPT 4.0 demonstrates a faster convergence toward consensus compared to the other examined LLMs. Specifically, it achieves consensus more frequently within the first round of proposal generation, even without user feedback, reaching 50 cases. This surpasses the number of immediate unanimous cases observed with Mistral Large 2 and AI21 Jamba, both of which remain below 40. When user feedback is incorporated, the differences among the three LLMs diminish over the first three iterations. Notably, Mistral Large 2 and AI21 Jamba require additional iterations to reach consensus in some cases (up to the fourth and fifth rounds), whereas ChatGPT 4.0 achieves consensus earlier.

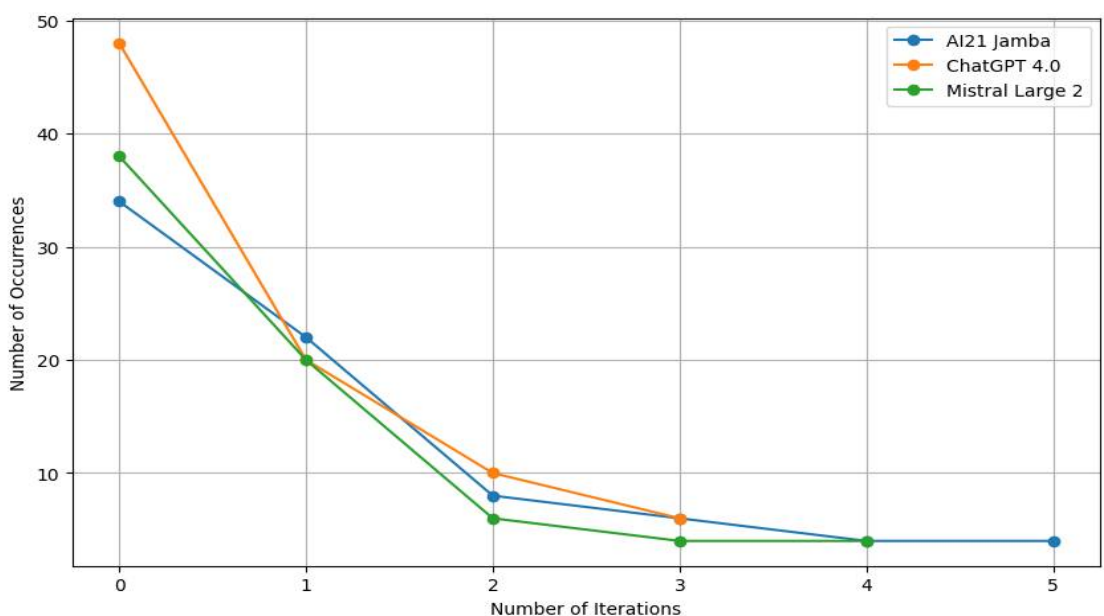

**Fig. 4**. Number of cases per iteration for achieving consensus

Lastly, aiming to understand how the similarity between states evolved throughout the discussions, an additional analysis focused on identifying the Elbow point [70] for each LLM by examining the average difference in Cosine Similarity from a value of 1 (fully identical) across iterations.

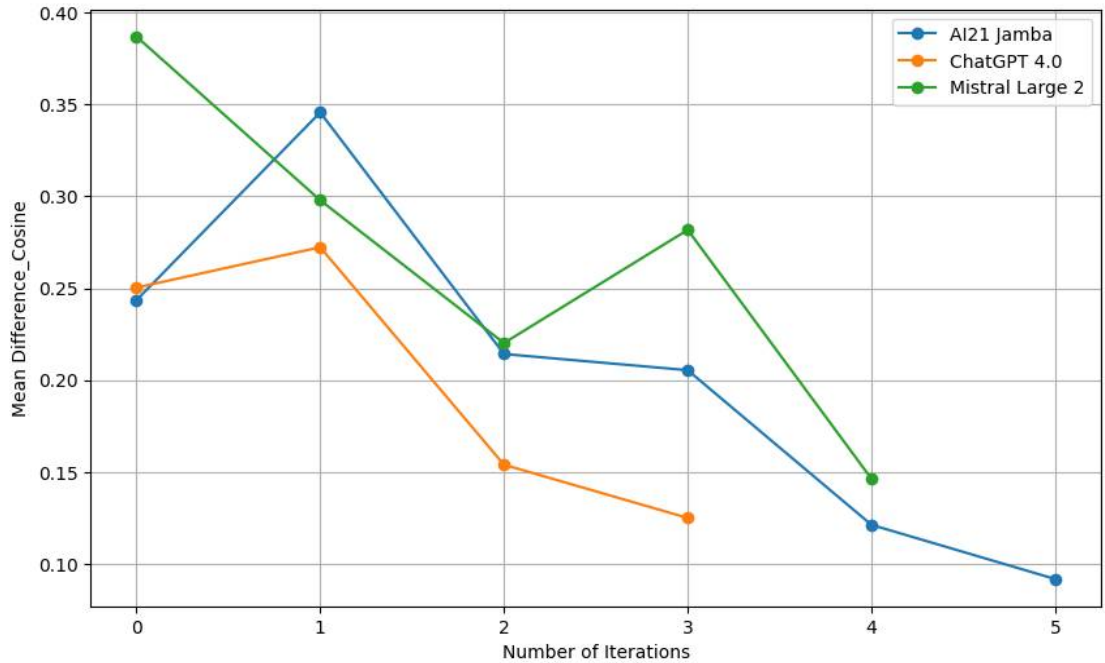

**Fig. 5**. Mean Difference in Cosine Similarity per Iteration

According to Figure 5, the data did not reveal a clear Elbow point for ChatGPT, Mistral Large 2, and AI21 Jamba. While ChatGPT 4.0 showed a steady decline, Mistral Large 2 and AI21 Jamba exhibited fluctuations, indicating that more complex discussions with additional iterations were likely needed for clarity.

## 6. Discussion

This study evaluated the potential of LLMs to facilitate consensus during controversial discussions, using cosine similarity as a key metric. By analyzing participants' initial positions and the consensus proposals ultimately accepted—generated by three widely used LLMs—valuable insights were obtained into the capacity of these innovative tools to contribute to this research area. Among the evaluated models—ChatGPT 4.0, Mistral Large 2, and AI21 Jamba —differences were observed in their approaches to generating consensus proposals. ChatGPT 4.0 consistently contributed to achieving consensus across all discussions on the SDGs by creating proposals that closely aligned with participants' initial positions. In contrast, while AI21 Jamba and Mistral Large 2 were equally effective in leading users to consensus, they tended to generate proposals that diverged more significantly from the participants' original views. This trend was consistent across all discussions in this pilot study, which addressed topics spanning the 17 SDGs. Notably, ChatGPT 4.0 demonstrated the greatest similarity to participants' initial positions in discussions on climate change, whereas AI21 Jamba adopted the most divergent approach in formulating consensus proposals on the same topic.

ChatGPT 4.0's strategy of aligning its proposals more closely with users' initial opinions appeared to play a significant role in achieving consensus more quickly. By evaluating the intermediate stages required to reach consensus, it was observed that ChatGPT 4.0 facilitated agreement after only one or two iterations of the application process. This indicates that it effectively selected and implemented the optimal strategy from the available options, a level of efficiency not demonstrated to the same extent by the other two LLMs, Mistral Large 2 and AI21 Jamba. Although the latter produced commendable results in achieving consensus, their proposals required more iterations to reach an acceptable outcome. Specifically, these proposals only aligned with users' opinions and feedback after 4 to 5 iterations, ultimately achieving consensus but with greater time and effort.

Despite the interest these findings have generated regarding the ability of LLMs to facilitate consensus, it is important to acknowledge some limitations of this study. First, while Cosine Similarity is a popular metric for assessing consensus facilitation by LLMs, it may not fully capture the qualitative aspects of participant agreement. Additionally, the fact that the agreement was achieved between only two participants, both students, may limit the generalizability of the findings to broader and more diverse populations. Similarly, the study's focus on discussions related to sustainable development issues may restrict the applicability of the results to other discussion contexts. Finally, the discussions in this study were conducted in Greek, which could have negatively influenced the performance of the models examined, given that the primary training languages for ChatGPT 4.0 and AI21 Jamba were English, while Mistral Large 2 was predominantly trained in French.

Given these limitations, future research could build on this study by incorporating more diverse and larger groups of participants to enhance the generalizability of the findings. Broadening the scope of discussion topics beyond sustainable development issues could also provide insights into the applicability of LLMs in facilitating consensus in various contexts. Furthermore, the possibility of using additional strategies, such as voting, structured debate, or collaborative brainstorming, could provide a deeper understanding of how different approaches influence consensus-building. And indeed, without limiting ourselves to cosine similarity, further metrics such as Jaccard similarity, semantic distance, or agreement indices could be tested to capture both quantitative and qualitative aspects of consensus. Finally, it would be important to implement this application scenario in the future in multiple languages, including those in which the models were primarily trained, to assess the impact of language on model performance and provide a more nuanced understanding of their capabilities in multilingual environments.

## 7. Conclusion

The present research demonstrates the transformative potential of LLMs in facilitating consensus-building, particularly in structured online discussions. ChatGPT 4.0 emerged as the most effective facilitator, consistently generating consensus proposals aligned with participants' initial views and achieving agreement within fewer iterations compared to Mistral Large 2 and AI21 Jamba. These results affirm the capability of LLMs to serve as neutral mediators in diverse and complex discussions, while identifying areas for improvement, such as addressing linguistic and cultural diversity. Future research should expand on these findings by exploring larger and more diverse participant groups, employing additional consensus strategies, and incorporating qualitative measures to capture the nuanced dynamics of agreement formation. These advancements can further refine the integration of AI into collective decision-making frameworks, fostering more inclusive, efficient, and sustainable collaboration.